\documentclass[3p,number,sort&compress,times]{elsarticle}

\usepackage{graphicx}
\usepackage{dcolumn}
\usepackage{bm}
\usepackage{amsmath}
\usepackage{amssymb}

\usepackage{hyperref}
\usepackage{upgreek}
\usepackage{color}

\begin{document}

\title{Networks of ABA and ABC stacked graphene on mica\\ observed by scanning tunneling microscopy}

\author{S. Hattendorf\corref{cor1}}
\ead{Hattendorf@physik.rwth-aachen.de}
\author{A. Georgi\corref{cor2}}
\author{M. Liebmann}
\author{M. Morgenstern}
\address{II. Physikalisches Institut B, RWTH Aachen and JARA-FIT, Otto-Blumenthal-Stra{\ss}e, 52074 Aachen}
\cortext[cor1]{Corresponding author}

\begin{abstract}
Graphene flakes are prepared on freshly cleaved mica by exfoliation and studied by scanning tunneling microscopy in ultra high vacuum.
On few-layer graphene, a triangular network of partial dislocations separating ABC stacked and ABA stacked graphene was found similar to the networks occasionally visible on freshly cleaved HOPG. We found differences in the electronic structure of ABC and ABA stacked areas by scanning tunneling spectroscopy, i.e., a pronounced peak at 0.25\,eV above the Fermi level exclusively in the ABA areas, which is shown to be responsible for the different apparent heights observed in STM images.
\end{abstract}



\maketitle

\section{\label{sec:Intro}Introduction}
 Since Geim and Novoselov succeeded in producing monolayer graphene flakes by exfoliation \cite{Novoselov2004}, graphene developed into a favorite research topic due to its two-dimensional nature and its unique combination of electronic, mechanical and optical properties accompanied by chemical inertness \cite{Castro-Neto,Novoselov2007}.

\begin{figure}
\centering
\includegraphics[width=0.3\textwidth]{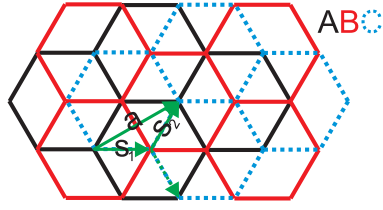}
\caption{\label{fig:Burger}Sketch of ABC stacked trilayer graphene with the three layers colored differently as marked. Only the carbon-carbon bonds are shown as lines. The arrow labeled 'a' marks a Burgers vector of a possible dislocation in the upper graphene layer separating two areas of the same stacking.
This dislocation can be separated into two shorter (and thus energetically more favorable) partial dislocations (labeled 's$_1$' and 's$_2$'). Since 's$_1$', for example, transforms an A-layer into a C-layer, this results in adjacent, differently stacked areas, i.e. ABC and CBC. The third possible direction of a partial dislocation is shown as a dashed arrow.}
\end{figure}

Recently, trilayer graphene, came into focus \cite{Craciun2011,Bao2011,Zhang2011a,Lui2011}. For more than two layers, graphene can be stacked either in Bernal-stacking showing an ABA sequence or in rhombohedral stacking with an ABC sequence (s. Fig.\ \ref{fig:Burger}). The latter is energetically slightly less favorable by about 0.18 \,meV per atom \cite{Aoki2007}. For both stackings, the distances between carbon atoms within the graphene sheet (1.418\,$\rm{\AA}$) and the separations of graphene layers (3.348\,$\rm{\AA}$) are identical \cite{Lipson1942}. The lateral energy barrier between these two stackings has been calculated to be about 2 meV per atom only, which favors a network of stacking faults \cite{Aoki2007}.
Networks of partial dislocations connecting triangular ABA and ABC areas have indeed been observed in highly oriented pyrolytic graphite (HOPG) by electron microscopy \cite{Williamson1960, Amelinckx1960, Delavignette1960, Delavignette1962}. The partial dislocation character has been proven directly in dark-field images \cite{Williamson1960}.
STM images of such partial dislocation networks were recorded subsequently
revealing an apparent height contrast between differently stacked areas \cite{Snyder1992, Snyder1993, Ouseph1996, Kobayashi2004, Kobayashi2004a}. Kobayashi et.\,al. \cite{Kobayashi2004a} showed that this contrast slightly changed in intensity across a step edge, which was tentatively explained by a difference in the local density of states (LDOS) resulting from an interference of electronic waves normal to the surface.

The stacking order, moreover, influences the electrical properties of graphene. It was predicted that an electric field $\underline{E}$ can open a band gap of up to 200 meV for ABC (rhombohedral) stacking, but not for the conventional ABA (Bernal) stacking \cite{Guinea2006,Latil2006,Lu2006, Lu2007,Aoki2007,Kumar2011a,Avetisyan2010,Tang2011,Xiao2011}. Thicker layers of ABC stacking exhibit conducting surface states in the absence of $\underline{E}$, but an insulating bulk, very similar to weak topological insulators \cite{Xiao2011}. In addition, high-temperature superconductivity driven by the large density of states at the surface has been proposed for thicker ABC stacked layers \cite{Kopnin2011}. 
The two different stackings of trilayers have been distinguished experimentally by gate dependent infrared spectroscopy \cite{Mak2010,Lui2011}, by Raman spectroscopy \cite{Lui2011_2,Jhang2011}, and by differences in the temperature- or magnetic-field dependent transport properties \cite{Bao2011,Zhang2011a,Kumar2011,Taycha2011}. This revealed, e.g., that about 15 \% of the sample area prepared by exfoliation is ABC stacked \cite{Lui2011_2}, and that the predicted distinction in band-gap opening exists, although so far only with a band gap below 10 meV \cite{Jhang2011,Bao2011,Khodkov2012}. These methods, however, are limited in lateral resolution to about 1 $\mu$m, thus, they are not able to resolve finer stacking orders in graphene.

Here, we show that such finer networks of ABA and ABC stacked graphene exist. In particular, graphene on mica exhibits a network of ABA and ABC stacked areas with sizes down to (200\,nm)$^2$ very similar to freshly cleaved HOPG \cite{Snyder1992, Snyder1993, Ouseph1996, Kobayashi2004, Kobayashi2004a}. We further identify a peak at 200\,meV by scanning tunneling spectroscopy, which is only visible in the ABA stacked region and turns out to be responsible for the different apparent height in STM images. This peak, thus, makes the distinction of the different stackings within the network possible.

\section{\label{sec:Ex}Experiment}

Graphene samples were prepared by mechanical exfoliation \cite{Novoselov2004} on freshly cleaved muscovite mica (highest quality "V1", Plano GmbH, Wetzlar, Germany), an aluminosilicate with the formula KAl$_2$(Si$_3$,Al)O$_{10}$(OH)$_2$ \cite{Rudenko2011} and a layered structure consisting of one layer of octahedrally coordinated Al$^{3+}$ions surrounded by two layers of tetrahedral Si$^{4+}$ ions. Every fourth Si$^{4+}$ ion is replaced by an Al$^{3+}$, resulting in an excess negative charge, which is compensated by K$^{+}$ ions connecting the triple layers. Mica can be cleaved along the K$^{+}$ layers with large areas (several 100 $\mu$m$^2$) of mono-atomic flatness.
However, the density distribution of K$^+$ on the two cleaved surfaces is not perfectly homogeneous leading to hydrophilic properties
which trigger the formation of potassium carbonate islands on the surface, if ambient water is present \cite{Ostendorf2008, Ostendorf2009}.

In order to reduce the influence of water, we cleaved mica and deposited graphene on top by exfoliation within a dry air box with reduced, though not explicitly controlled, water density. We did not find any differences in the amount of elevations previously attributed to subsurface H$_2$O \cite{Xu2010,Shim2012,He2012} using exfoliation in air, in a dry box or in an argon box with humidity below 2\,ppm.

Thin graphene flakes were identified afterwards using optical polarization microscopy under ambient conditions \cite{Dorn2010}. The number of layers was determined by Raman spectroscopy \cite{Malard2009} at laser excitation wavelength of 532\,nm.
To avoid extensive heating of the sample, the laser power was kept below 2\,mW and the measurement time was 5-10\,s. To determine the thickness of few-layer flakes, tapping mode AFM was used to measure the height difference between a monolayer area of graphene and the multilayer area of interest. The surface roughness is determined by STM to be $(61\pm13)$\,pm on the monolayer areas and $(23\pm2)$\,pm on multilayer areas, slightly higher than in \cite{Lui2009, Xu2010}, but significantly lower than for graphene on SiO$_2$ \cite{Zhang2008,Geringer2009,Ryu2010}.
To avoid possible chemical residues on the graphene surface resulting from lithography \cite{Geringer2010}, indium micro-soldering was used for contacts \cite{Girit2007}. This required heating of the sample to 180-190$^{\circ} \rm{C}$ in order to melt the In onto the graphene surface.

The STM measurements were carried out at 300 K in a home-built UHV system operating at a background pressure of 1-2$\cdot 10 ^{-8}\,\rm{Pa}$. Tungsten tips were prepared by electrochemical etching and subsequent short heating inside the ultrahigh vacuum chamber to remove the tungsten trioxide layer formed during the etching process \cite{Lucier2005}.
The positioning of these tips above the graphene flakes is achieved using a long-range optical microscope with resolution down to 10 $\mu$m \cite{Geringer2009}.
A bias voltage $V$ of $V=0.2$ - $0.6\,\rm{V}$ or $V=-0.2$ - $-0.5\,\rm{V}$ was applied to the graphene sample and currents of $I=0.05$ - $0.5\,\rm{nA}$ were used for constant-current images. The $dI/dV$ curves were acquired by applying an additional ac-voltage with a frequency of 1.43\,kHz and a modulation amplitude of $V_{\rm mod}=20-40\,\rm{mV}$, corresponding to an energy resolution of $97-130\,\rm{meV}$, and detecting the resulting in-phase ac current by a lock-in amplifier.

\section{\label{sec:RAD}Results and Discussion}

\subsection{\label{sec:Pre}Raman spectroscopy and atomic force microscopy}

\begin{figure}
\centering
\includegraphics[width=0.48\textwidth]{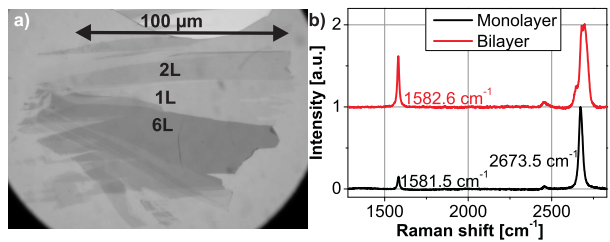}
\caption{\label{fig:Flocken}(a) Optical image of a sample prepared by exfoliation in the dry box recorded with polarization microscopy; layer thickness $n$ in monolayers of graphene is marked as $n$L. (b) Corresponding Raman spectra of the regions marked in the optical images as 1L (Monolayer) and 2L (Bilayer) with wave numbers corresponding to the peaks indicated.}
\end{figure}
Figure \ref{fig:Flocken} shows an optical image of the prepared sample along with the corresponding Raman spectra. Only this sample (one of three) exhibited the triangular pattern within the area labeled '6L'. The number of layers is identified as usual by analyzing the shape and position of the 2D peak \cite{Malard2009}. The height of 6 layers was determined by AFM.

\subsection{\label{sec:Sta}Stacking-faults in few-layer graphene on mica}
\begin{figure*}
\includegraphics[width=1\textwidth]{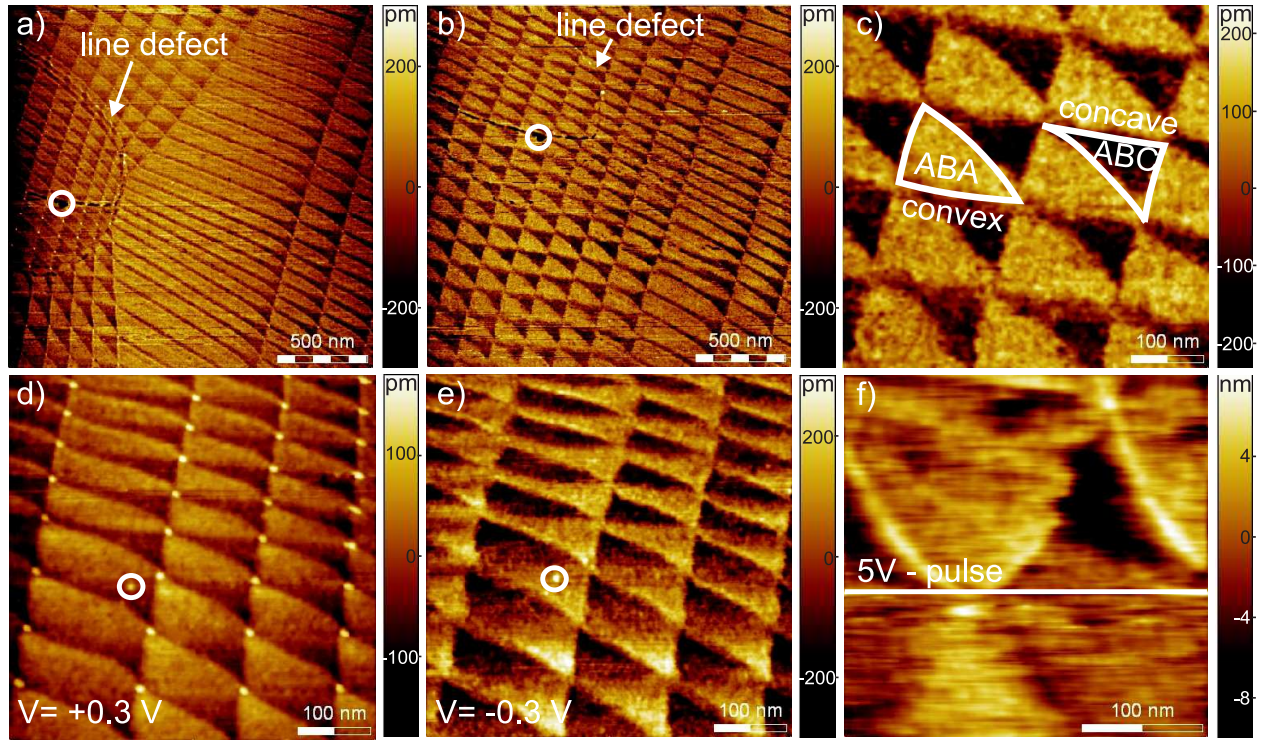}
\caption{\label{fig:Dreiecke}STM images of partial dislocation network on few-layer (6 $\pm$ 1 layers) area of graphene on mica. (a), (b)  are recorded on the same area with 2 hours in between; $V=0.4$\,V, $I=0.1$\,nA, scan speed: 2 $\frac{\upmu\rm{m}}{\rm{s}}$;  white circles highlight a defect which marks the same area in (a) and (b); a line defect is indicated in both images; notice the change of the network towards a more regular structure from (a) to (b);
(c) Zoom into the lower left edge of (b); white lines mark partial dislocation lines, which according to the indicated curvature presumably encircle ABA and ABC areas as marked; (d), (e) are recorded subsequently on the same area with different voltage polarity; white circles mark a defect, i.e. the same position in both images; (d) $V=0.3$\,V; $I=0.15$\,nA; 1 $\frac{\upmu\rm{m}}{\rm{s}}$; (e) $V=-0.3$\,V; $I=0.15$\,nA; 1 $\frac{\upmu\rm{m}}{\rm{s}}$; notice the contrast inversion between (d) and (e). (f) STM image recorded at $V=0.4$ V; $I=0.2$ nA; 0.6 $\frac{\upmu\rm{m}}{\rm{s}}$; during scanning, a 5\,V pulse is applied at the white line as marked, leading to contrast inversion.}
\end{figure*}

STM images of the few-layer graphene area, shown optically in Fig.\ \ref{fig:Flocken} (a), are displayed in Fig.\ \ref{fig:Dreiecke} revealing an area with a triangular pattern. Triangles pointing in two different directions exhibit apparent heights differing by 1-4\,\AA .
Due to the similarity of theses patterns to the patterns found previously on HOPG \cite{Williamson1960, Amelinckx1960, Delavignette1960, Delavignette1962,Snyder1992, Snyder1993, Ouseph1996, Kobayashi2004, Kobayashi2004a},
we assign them to a partial dislocation network separating triangular areas of
ABA and ABC stacked graphene. 
Fig.\ \ref{fig:Dreiecke} (a) and (b) show two large-scale images, which have been measured subsequently on the same area of the sample. Obvious changes in the pattern towards a more regular network emphasize the large mobility of the partial dislocation network, which is a well-known property of partial dislocations \cite{Bacon2009}.
Within the first image, some partial dislocation lines appear to be pinned to the defect line marked in Fig.\ \ref{fig:Dreiecke} (a),  but are depinned two hours later in Fig.\ \ref{fig:Dreiecke} (b). We checked carefully that the appearance of such a relaxed structure does not depend on scan direction, which, however, does not exclude that the depinning was initiated by the scanning process.
The large scale mobility of the network underlines that the apparent height is not caused by a topographic feature, which is unlikely to be moved strongly at 300 K once and then remain stable.
The triangular shape of the pattern is, in addition, in line with the three possible directions of Burgers vectors transforming ABA into ABC areas (s. Fig. \ref{fig:Burger}).
In Fig.\ \ref{fig:Dreiecke} (c), the shape of a dark and a bright region is marked by white lines. One observes that the bright one has convex edges, while the dark one exhibits concave edges. Based on this shape, the stacking order of the different areas can be assigned. Since ABC stacking is energetically less favorable \cite{Aoki2007}, there is a force on the partial dislocation lines in the direction of ABC stacking, resulting in concave shapes. In turn, this identifies the darker areas observed at positive sample bias as ABC regions or, more generally, as the stacking fault regions.
The crossing points of the partial dislocation lines, called nodes, require an energetically costly local AA stacking or a vacancy due to symmetry \cite{Hohage1995}. These nodes often appear brighter than the surrounding area (e.g Fig.\ \ref{fig:Dreiecke} (d)).

The contrast between ABA and ABC stacked areas as shown in Fig.\ \ref{fig:Dreiecke}(d) and (e), inverts with bias polarity. This was not observed on HOPG, where only a gradual change was found \cite{Snyder1992}. Moreover, the contrast could be inverted by voltage pulses changing the shape of the tip. This is shown in Fig.\ \ref{fig:Dreiecke}(f) and verified by imaging the complete area prior and after the voltage pulse with the same tunneling voltage of 0.4 V (not shown).
Since the tip, where we occasionally observed the contrast change by pulsing, was rather unstable as visible in the STM image and exhibited
an unusually large decay length in $I(z)$ curves, we believe that
the tip chemistry and, thus, the surface potential of the tip changed significantly during the pulse. Importantly, all these observations strongly suggest that the observed contrast is not of topographic origin but of an electronic one.

\begin{figure}
\centering
\includegraphics[width=0.48\textwidth]{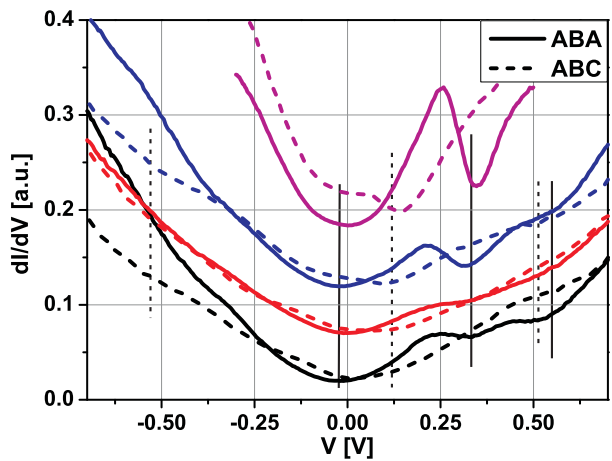}
\caption{\label{fig:dIdU}$dI/dV$ spectra acquired on ABA stacked (solid) and ABC stacked (dashed) regions. Spectra of the same color were acquired on adjacent regions directly after each other. The black, red and blue measurements are stabilized at $V=1$\,V and $I=0.2$\,nA, $V_{\rm mod}=40$ meV, the pink ones at $V=0.5$\,V and $I=0.2$\,nA, $V_{\rm mod}=20$ meV. Measurements are shifted vertically for clarity. Vertical lines mark minima.}
\end{figure}

To further analyze the electronic differences, $dI/dV$ spectra, which are proportional to the local density of states (LDOS) \cite{Morgenstern2000,Tersoff1985}, were acquired on ABC and ABA areas as shown in  Fig.\ \ref{fig:dIdU}.
The spectra exhibit several minima marked by vertical lines and an obvious peak at $V=0.25\,\rm{V}$ only on the ABA areas. Also the positions of the minima are different for differently stacked areas as marked. Since the curves were acquired alternately on ABA and ABC stacked areas, one can exclude that the spectroscopic differences are caused by a tip change.
The global minimum is found around 0\,mV for ABA areas and at 125\,mV for ABC areas. This might indicate a relative offset of the Dirac point energy $E_{\rm D}$. The peaks and dips are slightly moved upwards by about 20 mV in the spectra acquired at smaller tip-surface distance (pink curves). This indicates an upwards band bending at the graphene surface induced by the work function mismatch between tip and graphene and implies an electric field penetration into the graphene sample \cite{Morgenstern2000}.
No band gap is observed neither on the ABA nor on the ABC areas, which means that a band gap, if present, is smaller than about 100 meV being the energy resolution of the experiment \cite{Morgenstern2000}.
Theoretical calculations \cite{Lu2006, Lu2007,Xiao2011} of ABA and ABC stacked graphene find complex band structures changing with electric field, which,
however, are largely symmetric with respect to $E_{\rm D}$. We checked that they cannot explain a single peak at +0.25 eV, which is only found in ABA areas, using reasonable assumptions for the electric field penetration, which itself depends on tip-sample distance and screening. Since the peak is observed close to the tentative $E_{\rm D}$ of the ABC area, i.e. where a gap exists in the bulk of the ABC stacked areas, it might indicate a confinement of electronic states within the ABA areas (see below).

\begin{figure}
\centering
\includegraphics[width=0.48\textwidth]{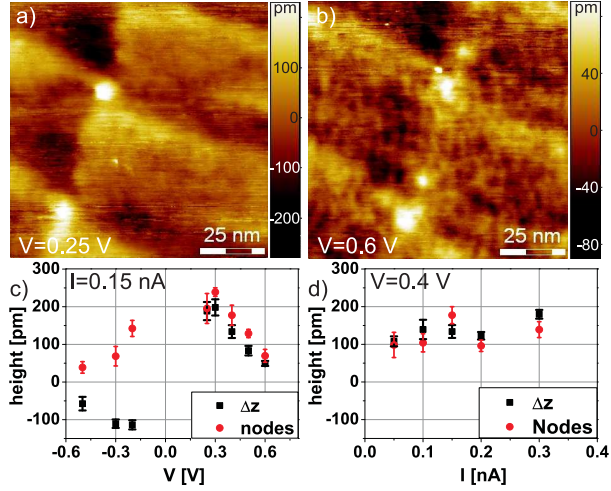}
\caption{\label{fig:Hoehenprofs}(a) Small scale STM image of multilayer graphene with stacking faults, $V=0.25\,\rm{V}$, $I=0.15\,\rm{nA}$; (b) same area as in (a), $V=0.6\,\rm{V}$; $I=0.15\,\rm{nA}$.  (c) Voltage dependency of the averaged height difference between ABA and ABC areas $\Delta z$ and of the average height of the nodes with respect to the highest stacking area in the surrounding (nodes); $I= 0.15\,\rm{nA}$. (d) Current dependency of the same quantities as displayed in (c); $V=0.4\,\rm{V}$.}
\end{figure}
Next, we check that the spectroscopic data largely explain the apparent height difference between ABA and ABC stacked graphene. This excludes a mechanical influence of the electric field on the tip of the graphene \cite{Mashoff2010} to be different on different stacking areas.
Therefore, we first measure the dependency of $\Delta z$ on sample voltage $V$ and current $I$. Figure \ref{fig:Hoehenprofs}(a) and (b) show part of the partial dislocation network imaged at two different $V$, where the height difference between the darker areas (ABC) on the left and the brighter areas (ABA) on the right has been used to measure $\Delta z$. The results are displayed in Fig.\ \ref{fig:Hoehenprofs}(c) and (d). In addition, the height of the nodal points is determined with respect to the surrounding area appearing the highest and is shown as 'nodes' in Fig.\ \ref{fig:Hoehenprofs}(c) and (d), too.
Obviously, the current dependency of $\Delta z$ is weak while the voltage dependency is pronounced indicating that the distance between tip and sample plays a minor role. The voltage dependent contrast inverts with voltage polarity, increases with decreasing $|V|$, and is largest around $V=0.25$ V, close to the position of the peak in $dI/dV$ curves. Similar trends are observed for the contrast of the nodes, except that the nodes always appear higher than the surrounding. Partly, the partial dislocation lines exhibit a distinct contrast as visible in Fig.\ \ref{fig:Hoehenprofs}(b), but we did not evaluate this contrast in more detail, because it appears only at certain $V$.

\begin{figure*}
\includegraphics[width=1\textwidth]{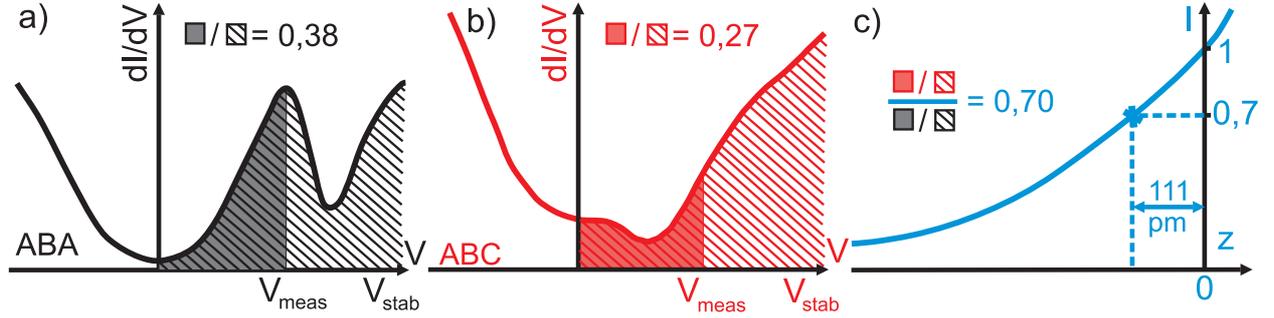}
\caption{\label{fig:Skizze-HSim}Illustration on how to determine the expected height difference between ABA and ABC stacked areas from $dI/dV$ spectra; (a),(b) Lines: $dI/dV$ spectra of ABA  and ABC stacked areas as marked (same as pink curves in Fig.\ \ref{fig:dIdU}); the hatched area up to $V_{\rm stab}$ has to be integrated to determine the current relevant for the stabilization during the $dI/dV$ curve, while the completely filled area up to $V_{\rm meas}$ determines the current at $V_{\rm meas}$; the ratio between filled and hatched areas is given above the curves; (c) sketch of the distance dependence of the current; the ratio of the ratios from (a) and (b), 0.7, is marked symbolically and at the $I$-axis; it is translated to a height difference by the $I(z)$ curve as marked.}
\end{figure*}

\begin{figure}
\centering
\includegraphics[width=0.48\textwidth]{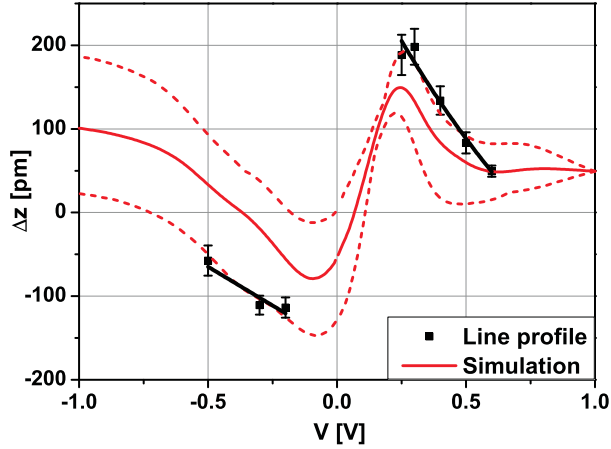}
\caption{\label{fig:dIdUHoehenSim}Black Squares: height difference $\Delta z$ between ABA and ABC stacked areas measured by evaluating a line profile in constant-current images (same data as in Fig.\ \ref{fig:Hoehenprofs}(c)); black solid lines serve as guide-to-the-eye; red solid line: averaged height differences calculated from five $dI/dV$ spectra stabilized at $V_{\rm stab}=1\,\rm{V}$ according to the method illustrated in Fig.\ \ref{fig:Skizze-HSim}; red dashed lines: error bars resulting from fluctuations for $dI/dV$ spectra taken at different positions on the sample.}
\end{figure}

Now, we calculate the apparent height difference expected from the $dI/dV$ curves.
According to Tersoff and Hamann \cite{Tersoff1985}, the tunneling current $I$ at voltage $V$ depends on the LDOS of the tip $\rho^{\rm t}(E)$ and the LDOS of the sample at the position below the tip $\rho^{\rm s}(E)$ according to:
\begin{equation}
I\propto\int_0^{eV}\rho^{\rm t}\left(E_{\rm F}+\epsilon\right)\cdot \rho^{\rm s}\left(E_{\rm F}-eV+\epsilon\right)\cdot T(\epsilon,V,z)d\epsilon.
\label{eq:Iwacho}
\end{equation}
Here, $E_{\rm F}$ is the Fermi level of the tip, $z$ is the distance between the tip and the sample surface, e is the electron charge, and $T(\epsilon,V,z)$ is the transmission coefficient of the tunneling process being exponentially dependent on $z$.
Since a constant current is used to stabilize the vertical tip position with respect to the sample, a difference in LDOS of the different sample areas results in a change of the height difference $\Delta z$ with voltage. To estimate the expected $\Delta z(V)$, the $dI/dV$ curves are integrated between 0\,V and the applied voltage $V_{\rm meas}$ and translated into a height difference using $I(z)$ spectra. The procedure is illustrated in Fig.\ \ref{fig:Skizze-HSim} exemplary for the pink curves of Fig.\ \ref{fig:dIdU}. As can be seen by comparing Fig.\ \ref{fig:Skizze-HSim}(a) and (b), the $dI/dV$ values in the completely filled area of the curve are larger for ABA stacked graphene than for ABC stacked graphene.
Thus, the resulting height difference must increase by reducing the voltage from $V_{\rm{stab}}$ to $V_{\rm{meas}}$. This is indeed observed experimentally.
To determine the expected $\Delta z$ quantitatively, the filled area underneath the $dI/dV$ curves, ranging from $V=0\,\rm{V}$ to $V_{\rm meas}$, is divided firstly by the hatched area, ranging from $V=0\,\rm{V}$ to $V_{\rm stab}$.
Then, the resulting ratios obtained for ABA and ABC area are divided by each other and are used to determine an additional height difference $\Delta z$ from the exponential $I(z)$ curve as illustrated in Fig.\ \ref{fig:Skizze-HSim}(c). For the $V_{\rm meas}$ in the example, an additional height difference of 111\,pm results.
Notice that the exponential $I(z)$ curve guarantees an independence of the found $\Delta z$ from the chosen $z$ value in the $I(z)$ curve.\\
Figure \ref{fig:dIdUHoehenSim} shows the calculated $\Delta z (V_{\rm meas})$ using $dI/dV$ curves stabilized at $V_{\rm stab}=1\,\rm{V}$ (red line) in comparison with the measured $\Delta z (V)$ from Fig.\ \ref{fig:Hoehenprofs}(c). Dashed lines mark the error bar determined from five pairs of $dI/dV$ curves recorded at different positions. As the topographic data at $V=1\,\rm{V}$ became rather noisy, the height difference at $V=0.6\,\rm{V}$ was added as the offset to determine the calculated heights. Note that the simulated height difference changes very little between $V=0.6$ and $V=1\,\rm{V}$.
The characteristics of the measured $\Delta z(V)$ curve, i.e., the change of contrast around $V=0\,\rm{V}$ and the decrease of contrast with increasing $|V|$, are well reproduced by the calculation, although the experimental $|\Delta z|$ values are slightly larger. The reasonable agreement excludes a dominating influence of a mechanical interaction between the tip and the graphene as well as of the electric field of the tip onto the apparent height differences, since both are changing with distance being significantly different in the constant-current image used to determine the experimental
$\Delta z$ and in the $dI/dV$ curve used to determine the calculated $\Delta z$. This is in line with the observation that $\Delta z$ barely changes with $I$.
Since the contrast inversion with voltage polarity excludes a dominating influence of topography, we conclude that the different LDOS is primarily responsible for the contrast between ABA and ABC areas in constant-current images.

The question remains: what causes the differences in LDOS, respectively the LDOS peaks? Intrinsic differences of the band structure could be excluded as discussed above. A Fabry-Perot type interference, similar to a proposal by Kobayashi et.\,al. \cite{Kobayashi2004a}, was considered. The idea is the following: electrons tunneling into the graphene multi-layer are reflected back at the interface between graphene and mica forming an interference pattern with the incoming electrons. Depending on the energy of the electrons, this interference can be constructive or destructive causing peaks and minima in the LDOS. The energy of these peaks is straightforwardly calculated using the energy dispersion of graphite orthogonal to the surface \cite{Kobayashi2004a}:
\begin{equation}
E = \frac{(\hbar k)^2}{2m_{\perp}}-2\gamma_1
\label{eq:EnergyDispersion}
\end{equation}
with $m_{\perp}=\frac{\hbar^2}{2c^2\gamma_1}$, where $\hbar$ is the reduced Planck constant, the interlayer distance $c = 335\,\rm{pm}$ and the interlayer hopping parameter $\gamma_1 = 0.39\,\rm{eV}$ \cite{Holzwarth1982} as well as the condition for constructive interference for the wave vector $k$:
\begin{equation}
k = \frac{2\pi}{\lambda} = \frac{\pi n}{c N}
\label{eq:Interference}
\end{equation}
with $n \in \mathbb{N}$, $\lambda$ being the wavelength of the electrons, and $N$ being the number of layers in the graphene sheet.

The stacking fault might scatter the electrons isotropically due to the band gap in its interior prohibiting the vertical interference, and, thus, explaining the absence of strong peaks in the ABC region.  However, for the determined layer thickness $N=6$, the four lowest energy peaks would be at about -0.7 eV, -0.35 eV, +0.2 eV, +0.9 eV with respect to $E_{\rm D}$. This is in disagreement with the measured spectra exhibiting a single strong peak at about 0.25 eV and maybe a smaller one at 0.4 eV, but no strong peak around -0.35\,eV. Layer numbers $N$ between 5 and 7 have been checked, in accordance with the maximum uncertainty of the layer thickness determined by AFM, but without success.
Thus, such a model cannot explain the observed $dI/dV$ curves straightforwardly. We propose tentatively that an additional lateral confinement by the gap within the interior of the ABC area strengthens the vertical Fabry-Perot resonance in the ABA region which is closest to $E_{\rm D}$, i.e. the one at 0.2 eV.
Thus, the ABA area surrounded by forbidden areas for the electrons at 0.2\,eV acts as a cavity for the electron waves. But more work is required to substantiate such a hypothesis. Notice finally that the network of stacking faults could be, in principle, transformed into a triangular network of 200\,nm  quantum dots by applying a gate voltage to open the band gap in the ABC regions \cite{Guinea2006,Latil2006,Lu2006, Lu2007,Aoki2007,Kumar2011a,Avetisyan2010,Tang2011,Xiao2011}

\section{\label{sec:Con}Conclusion}
In conclusion, we have shown that multilayer graphene can exhibit a regular pattern of alternating ABA and ABC stacking with stacking areas as small as (200\,nm)$^2$.  STS measurements reveal that the STM contrast between these areas is caused by differences in the local density of states which is largely dominated by a peak around 0.25 eV. This peak is only found on ABA areas and might be related to a confinement effect resulting from the topological character of ABC graphene.

\section*{Acknowledgments}
We gratefully acknowledge helpful discussion with E. Obraztsova and financial support by the DFG project Mo858/11-1 and Mo858/8-2.

\section*{References}
\bibliography{MicaPaper}

\end{document}